\documentclass[12pt, preprint,preprintnumbers,nofootinbib, groupedaddress,superscriptaddress,amsmath,amssymb]{revtex4}
\usepackage{graphicx}
\usepackage{dcolumn}
\usepackage{bm}
\usepackage{amssymb}
\usepackage{amsmath}
\usepackage{epsfig}    
\usepackage{color}
\usepackage{hhline}

\def\be{\begin{equation}}
\def\ee{\end{equation}}
\newcommand{\bea}{\begin{eqnarray}}
\newcommand{\eea}{\end{eqnarray}}
\newcommand{\nn}{\nonumber}

\numberwithin{equation}{section}

\begin{document}
{\begin{flushright}{APCTP Pre2022 - 014}\end{flushright}}

\title{Lepton mass matrix from double covering of $A_4$ modular flavor symmetry}

\author{Hiroshi Okada}
\email{hiroshi.okada@apctp.org}
\affiliation{Asia Pacific Center for Theoretical Physics (APCTP) - Headquarters San 31, Hyoja-dong,
Nam-gu, Pohang 790-784, Korea}
\affiliation{Department of Physics, Pohang University of Science and Technology, Pohang 37673, Republic of Korea}

\author{Yuta Orikasa}
\email{Yuta.Orikasa@utef.cvut.cz}
\affiliation{Institute of Experimental and Applied Physics, 
Czech Technical University in Prague, 
Husova 240/5, 110 00 Prague 1, Czech Republic}

\date{\today}

\begin{abstract}
We study a double covering of modular $A_4$ flavor symmetry
in which we construct lepton models in cases of canonical seesaw and radiative seesaw models. Through $\chi$ square numerical analysis, we show some predictions for the cases, depending on normal and inverted hierarchies.
\end{abstract}
\maketitle
\newpage

\section{Introduction}
Even after the discovery of the standard model(SM) Higgs, we puzzle about a way to accommodate neutral particles such as the active tiny neutrinos and dark matter(DM) candidate. 
To describe the observed neutrino sector,
we might need a prescription about how to determine three mixings and the two mass square differences in addition to CP phases; Majorana and Dirac phases, which are not precisely measured yet.
Modular flavor symmetries are one of the most promising candidates to obtain predictive scenarios in the neutrino sector, since
these symmetries do not require many neutral bosons~\footnote{These bosons are called flavons and they are traditionally introduced to get a desired neutrino mass matrix.}
due to  a new degree of freedom{: ``modular weight"}.
Moreover, DM can be stable by applying this degree of freedom.
In fact, vast amounts of literature along this line of idea have been appeared after the original paper~\cite{Feruglio:2017spp}.
\footnote{Charged-lepton and neutrino sectors have been discussed in ref.~\cite{deAdelhartToorop:2011re} by embedding subgroups of various finite modular flavor symmetries.}
For example, 
the modular $A_4$ flavor symmetry has been discussed in refs.~\cite{Kobayashi:2021ajl, Feruglio:2017spp, Criado:2018thu, Kobayashi:2018scp, Okada:2018yrn, Nomura:2019jxj, Okada:2019uoy, deAnda:2018ecu, Novichkov:2018yse, Nomura:2019yft, Okada:2019mjf,Ding:2019zxk, Nomura:2019lnr,Kobayashi:2019xvz,Asaka:2019vev,Zhang:2019ngf, Gui-JunDing:2019wap,Kobayashi:2019gtp,Nomura:2019xsb, Wang:2019xbo,Okada:2020dmb,Okada:2020rjb, Behera:2020lpd, Behera:2020sfe, Nomura:2020opk, Nomura:2020cog, Asaka:2020tmo, Okada:2020ukr, Nagao:2020snm, Okada:2020brs, Yao:2020qyy, Chen:2021zty, Kashav:2021zir, Okada:2021qdf, deMedeirosVarzielas:2021pug, Nomura:2021yjb, Hutauruk:2020xtk, Ding:2021eva, Nagao:2021rio, king, Okada:2021aoi, Nomura:2021pld, Kobayashi:2021pav, Dasgupta:2021ggp, Liu:2021gwa, Nomura:2022hxs, Otsuka:2022rak, Kang:2022psa, Ishiguro:2022pde,Nomura:2022boj, Kobayashi:2022jvy}, 
$S_3$  in refs.~\cite{Kobayashi:2018vbk, Kobayashi:2018wkl, Kobayashi:2019rzp, Okada:2019xqk, Mishra:2020gxg, Du:2020ylx}, 
$S_4$  in refs.~\cite{Penedo:2018nmg, Novichkov:2018ovf, Kobayashi:2019mna, King:2019vhv, Okada:2019lzv, Criado:2019tzk,
Wang:2019ovr, Zhao:2021jxg, King:2021fhl, Ding:2021zbg, Zhang:2021olk, gui-jun, Nomura:2021ewm}, 
$A_5$ in refs.~\cite{Novichkov:2018nkm, Ding:2019xna,Criado:2019tzk},
double covering of $A_4$  in refs.~\cite{Liu:2019khw, Chen:2020udk, Li:2021buv}, 
double covering of $S_4$  in refs.~\cite{Novichkov:2020eep, Liu:2020akv},   and
double covering of $A_5$  in refs.~\cite{Wang:2020lxk, Yao:2020zml, Wang:2021mkw, Behera:2021eut}.
Ref.~\cite{Kikuchi:2022geu} discusses CP phase of quark mass matrices in modular flavor symmetric models at the fixed point of $\tau$. 
Soft-breaking terms on modular symmetry is discussed in ref.~\cite{Kikuchi:2022pkd}.
Other types of modular symmetries have also been proposed to understand masses, mixings, and phases of the SM in refs.~\cite{deMedeirosVarzielas:2019cyj, Kobayashi:2018bff,Kikuchi:2020nxn, Almumin:2021fbk, Ding:2021iqp, Feruglio:2021dte, Kikuchi:2021ogn, Novichkov:2021evw, Kikuchi:2021yog, Novichkov:2022wvg}.~\footnote{Here, we provide useful review references for beginners~\cite{Altarelli:2010gt, Ishimori:2010au, Ishimori:2012zz, Hernandez:2012ra, King:2013eh, King:2014nza, King:2017guk, Petcov:2017ggy, Kobayashi:2022moq}.}
Different applications to physics such as dark matter and origin {of} CP are found in refs.~\cite{Kobayashi:2021ajl, Nomura:2019jxj, Nomura:2019yft, Nomura:2019lnr, Okada:2019lzv, Baur:2019iai, Kobayashi:2019uyt, Novichkov:2019sqv,Baur:2019kwi, Kobayashi:2020hoc, Kobayashi:2020uaj, Ishiguro:2020nuf, Ishiguro:2021ccl, Tanimoto:2021ehw}.
Mathematical {studies} such as possible correction from K\"ahler potential, systematic analysis of the fixed points, 
and moduli stabilization are discussed in refs.~\cite{Chen:2019ewa, deMedeirosVarzielas:2020kji, Ishiguro:2020tmo, Abe:2020vmv, Novichkov:2022wvg}.
Recently, the authors of ref. \cite{Kikuchi:2022txy} proposed a scenario to derive four-dimensional modular flavor symmetric models from higher-dimensional theory on extra-dimensional spaces with the modular symmetry. 
It constrains modular weights and representations of fields and modular couplings in the four-dimensional effective field theory. 
Higher-dimensional operators in the SM effective field theory are also constrained in the higher-dimensional theory, in particular, the string theory \cite{Kobayashi:2021uam}. Non-perturbative effects relevant to neutrino masses are studied in the context of modular symmetry anomaly \cite{Kikuchi:2022bkn}.

In this paper, we apply double covering of modular $A_4$ symmetry; $T'$, to the lepton sector, and show several predictions in cases of canonical seesaw scenario and radiative seesaw scenario~\cite{Ma:2006km}.
Notice here that the mathematical part of $T'$ has seriously been studied by ref.~\cite{Liu:2019khw}, and they also demonstrate a prediction in case of inverted hierarchy based on the canonical seesaw model.

This paper is organized as follows.
In Sec.~\ref{sec:realization}, we review our model setup in the lepton sector, giving superpotential, charged-lepton mass matrix, Dirac Yukawa matrix, and the Majorana mass matrix.
In Sec.~\ref{neutrino},  
we formulate the neutrino mass matrix and their observables in case of canonical seesaw.
Then we move to the radiative seesaw model, showing the soft breaking terms that play a crucial role in generating the neutrino mass matrix at one-loop level. 
In Sec.~\ref{sec:num}, we perform the numerical $\chi$ square analysis, and show predictive figures in each cases of normal and inverted hierarchies of the canonical and radiative seesaws.
Finally, we conclude and summarize our model in Sec.~\ref{sec:conclusion}.
In Appendix A, we summarize formulas on the double covering of modular $A_4$ symmetry.

\begin{center} 
\begin{table}[tb]
\begin{tabular}{|c||c|c|c||c|c|c|c|c||}\hline\hline  
&\multicolumn{8}{c||}{ Chiral superfields}   \\\hline
  & ~$\{{\hat{L}_{e}},{\hat{L}_{\mu}},{\hat{L}_{\tau}}\}$~& ~$\{\hat{e}^c,\hat{\mu}^c,\hat{\tau}^c\}$~ & ~$\{\hat{N}^c_{1},\hat{N}^c_{2}\}$~ & ~$\hat{H}_1$~ & ~$\hat{H}_2$~& ~$\hat{\eta}_1$~ & ~$\hat{\eta}_2$~ & ~$\hat{\chi}$~
  \\\hline 
 $SU(2)_L$ & $\bm{2}$  & $\bm{1}$  & $\bm{1}$ & $\bm{2}$  & $\bm{2}$  & $\bm{2}$  & $\bm{2}$ & $\bm{1}$   \\\hline 
$U(1)_Y$ & $-\frac12$ & $1$ & $0$ & $\frac12$ & $-\frac12$& $\frac12$ & $-\frac12$  & $0$     \\\hline
 $T'$ & $3$ & $\{1,1'',1'\}$ & $2$   & $1$ & $1$ & $1$ & $1$ & $1$      \\\hline
 $-k$ & ${-1}$ & $-1$ & $-1$ & $0$ & $0$ & $-1$ & $-1$& $-3$ \\\hline
\end{tabular}
\caption{Field contents of {matter chiral superfields} and their charge assignments under $SU(2)_L\times U(1)_Y\times A_{4}$ in the lepton and boson sector, where $-k_I$ is the number of modular weight, and the quark sector is the same as the SM.}
\label{tab:fields}
\end{table}
\end{center}

\section{ Model} 
\label{sec:realization}
Here, we review our model in order to obtain the neutrino mass matrix.
In addition to the minimal supersymmetric SM (MSSM), we introduce matter superfields including two right-handed neutral fermions $N^c_{1,2}$ that belongs to doublet under {the modular} $T^\prime$ group with modular weight $-1$.
We also add three chiral superfields $\{\hat\chi, \hat{\eta}_1, \hat{\eta}_2 \}$ including two bosons $\{\chi, \eta_1, \eta_2\}$ where there superfields are true singlets under the $T^\prime$ group with $\{-1,-1,-3\}$ modular weight.
$\chi$ only plays a role in generating the neutrino mass matrix at one-loop level; therefore, $\eta_{1,2}$ are inert bosons as well as $\chi$. 
Left-handed lepton doublets $\{L_e,L_\mu,L_\tau\}$ are assigned to be triplet with $-1$ modular weight, while the right-handed ones $\{e^c,\mu^c,\tau^c\}$ to be $\{1,1'',1'\}$ with  $-1$ modular weight.
Two Higgs doublet $H_{1,2}$ are invariant under the modular $T^\prime$ symmetry.
All the fields and their assignments are summarized in Table~\ref{tab:fields}.
Under these symmetries, one writes renormalizable superpotential as follows~\footnote{Even though our assignments for each matter superfields are slightly different from the original paper in ref.~\cite{Liu:2019khw},  the resultant lepton mass matrix is same as the original one.}:
\begin{align}
&{\cal W} =
\alpha_e [Y^{(2)}_3 \otimes\hat{e}^c\otimes \hat{L} \otimes\hat{H}_2] 
+\beta_e [Y^{(2)}_3 \otimes \hat{\mu}^c\otimes \hat{L} \otimes\hat{H}_2] 
+\gamma_e [Y^{(2)}_3 \otimes \hat{\tau}^c\otimes \hat{L}\otimes \hat{H}_2]  
\nn\\
&+\alpha_\eta [Y^{(3)}_2 \otimes\hat{N}^c\otimes \hat{L} \otimes\hat{\eta}_1] 
+\beta_\eta [Y^{(3)}_{2''} \otimes\hat{N}^c\otimes \hat{L} \otimes\hat{\eta}_1] 
+ M_{0} [Y^{(2)}_3 \otimes\hat{N}^c\otimes\hat{N}^c] 
\label{eq:sp-lep}
\\
&+\mu_H \hat{H}_1 \hat{H}_2+\mu_\chi Y^{(6)}_1 \hat{\chi} \hat{\chi} 
 + a Y^{(4)}_1 \hat{H}_1 \hat{\eta}_2 \hat{\chi} + b Y^{(4)}_1 \hat{H}_2 \hat{\eta}_1 \hat{\chi}, 
\nn
\end{align}
where R-parity is implicitly imposed in the above superpotential, $Y^{(2)}_3\equiv(f_1,f_2,f_3)^T$ is $T'$ triplet with modular weight $2$, and $Y^{(3)}_{2^{(\prime\prime)}} \equiv(y^{(\prime\prime)}_1,y^{(\prime\prime)}_2)^T$ is $T^\prime$ doublet with modular weight $3$.~\footnote{The concrete expressions of modular Yukawas are summarized in Appendix A.} 
The first line in Eq.~(\ref{eq:sp-lep}) corresponds to the charged-lepton sector, while the second and third lines are related to the neutrino sector. Especially, the third line is important if the neutrino mass matrix is induced at one-loop level as dominant contribution.\\

After the electroweak spontaneous symmetry breaking,  the charged-lepton mass matrix is given by
\begin{align}
m_\ell&= \frac {v_2}{\sqrt{2}}
\left[\begin{array}{ccc}
\alpha_e & 0 & 0 \\ 
0 &\beta_e & 0 \\ 
0 & 0 &\gamma_e \\ 
 \end{array}\right]
 \left[\begin{array}{ccc}
f_1 &f_3 & f_2 \\ 
f_2 &f_1 & f_3 \\ 
f_3 &f_2 & f_1 \\ 
 \end{array}\right], 
\end{align}
where $\langle H_2\rangle\equiv [v_2/\sqrt2,0]^T$.
Then the charged-lepton mass eigenstate is found as ${\rm  diag}( |m_e|^2, |m_\mu|^2, |m_\tau|^2)\equiv V_{e_L}^\dag m^\dag_\ell m_\ell V_{e_L}$.
In our numerical analysis, we fix the free parameters $\alpha_e,\beta_e,\gamma_e$ inserting the observed three charged-lepton masses by applying the relations:
\begin{align}
&{\rm Tr}[m_\ell {m_\ell}^\dag] = |m_e|^2 + |m_\mu|^2 + |m_\tau|^2,\\
&{\rm Det}[m_\ell {m_\ell}^\dag] = |m_e|^2  |m_\mu|^2  |m_\tau|^2,\\
&({\rm Tr}[m_\ell {m_\ell}^\dag])^2 -{\rm Tr}[(m_\ell {m_\ell}^\dag)^2] =2( |m_e|^2  |m_\mu|^2 + |m_\mu|^2  |m_\tau|^2+ |m_e|^2  |m_\tau|^2 ).
\end{align}

The Dirac  matrix that consists of $\alpha_\eta$ and $\beta_\eta$; $N^c y_\eta L\eta_1$, is given by
\begin{align}
y_\eta &=
 \left[\begin{array}{ccc}
\frac{\beta_\eta}{\sqrt2} e^{\frac{7\pi}{12}i}y^{\prime\prime}_2 & \alpha_\eta y_1 & \frac{\alpha_\eta}{\sqrt2} e^{\frac{7\pi}{12}i}y_2
+ \beta_\eta y^{\prime\prime}_1 \\ 
 \frac{\beta_\eta}{\sqrt2} e^{\frac{7\pi}{12}i}y^{\prime\prime}_1
+ \alpha_\eta e^{\frac{\pi}{6}i} y_2 & \beta_\eta  e^{\frac{\pi}{6}i} y^{\prime\prime}_2 
& \frac{\alpha_\eta}{\sqrt2} e^{\frac{7\pi}{12}i}y_1  \\  \end{array}\right].
\end{align}
The heavier Majorana mass matrix is given by
\begin{align}
M_N &= M_0
 \left[\begin{array}{ccc}
f_2 & \frac1{\sqrt2} e^{\frac{7\pi}{12}i} f_3  \\ 
 \frac1{\sqrt2} e^{\frac{7\pi}{12}i} f_3 & e^{\frac{\pi}{6}i} f_1  \\
\end{array}\right]
=
M_0 {\tilde M}.
\end{align}
The heavy Majorana mass matrix is diagonalized by  a unitary matrix $V_N$ as follows: $D_N\equiv V_N M_N V_N^T$,
where $N^c\equiv \psi^cV_N^T$, $\psi^c$ being mass eigenstate.

\section{Active neutrino mass matrix}
\label{neutrino}
\subsection{Canonical seesaw}
If all the bosons have nonzero VEVs, the neutrino mass matrix is generated via tree-level as follows:
\begin{align}
m_\nu 
= \frac{v_{\eta_1}^2}{2 M_0} y_\eta^T \tilde M^{-1} y_\eta \equiv \kappa \tilde m_\nu,
\end{align}
where $\kappa\equiv  \frac{v_{\eta_1}^2}{2 M_0}$, and $\langle\eta_1\rangle \equiv [0,v_{\eta_1}/\sqrt2]^T$.
$m_\nu$ is diagonalzied by a unitary matrix $V_{\nu}$; $D_\nu=|\kappa| \tilde D_\nu= V_{\nu}^T m_\nu V_{\nu}=|\kappa| V_{\nu}^T \tilde m_\nu V_{\nu}$.
Then $|\kappa|$ is determined by
\begin{align}
(\mathrm{NH}):\  |\kappa|^2= \frac{|\Delta m_{\rm atm}^2|}{\tilde D_{\nu_3}^2-\tilde D_{\nu_1}^2},
\quad
(\mathrm{IH}):\  |\kappa|^2= \frac{|\Delta m_{\rm atm}^2|}{\tilde D_{\nu_2}^2-\tilde D_{\nu_3}^2},
 \end{align}
where $\Delta m_{\rm atm}^2$ is atmospheric neutrino mass difference squares, and NH and IH represent the normal hierarchy and the inverted hierarchy cases. 
Subsequently, the solar mass different squares can be written in terms of $|\kappa|$ as follows:
\begin{align}
\Delta m_{\rm sol}^2=  |\kappa|^2 ({\tilde D_{\nu_2}^2-\tilde D_{\nu_1}^2}),
 \end{align}
 which can be compared to the observed value.
 %
The observed mixing matrix is defined by $U=V^\dag_L V_\nu$~\cite{Maki:1962mu}, where
it is parametrized by three mixing angles $\theta_{ij} (i,j=1,2,3; i < j)$, one CP violating Dirac phase $\delta_{CP}$,
and one Majorana phase $\alpha_{21}$ as follows:
\begin{equation}
U = 
\begin{pmatrix} c_{12} c_{13} & s_{12} c_{13} & s_{13} e^{-i \delta_{CP}} \\ 
-s_{12} c_{23} - c_{12} s_{23} s_{13} e^{i \delta_{CP}} & c_{12} c_{23} - s_{12} s_{23} s_{13} e^{i \delta_{CP}} & s_{23} c_{13} \\
s_{12} s_{23} - c_{12} c_{23} s_{13} e^{i \delta_{CP}} & -c_{12} s_{23} - s_{12} c_{23} s_{13} e^{i \delta_{CP}} & c_{23} c_{13} 
\end{pmatrix}
\begin{pmatrix} 1 & 0 & 0 \\ 0 & e^{i \frac{\alpha_{21}}{2}} & 0 \\ 0 & 0 & 1 \end{pmatrix},
\end{equation}
where $c_{ij}$ and $s_{ij}$ stand for $\cos \theta_{ij}$ and $\sin \theta_{ij}$, respectively. 
Then, each of the mixings is given in terms of the component of $U$ as follows:
\begin{align}
\sin^2\theta_{13}=|U_{e3}|^2,\quad 
\sin^2\theta_{23}=\frac{|U_{\mu3}|^2}{1-|U_{e3}|^2},\quad 
\sin^2\theta_{12}=\frac{|U_{e2}|^2}{1-|U_{e3}|^2}.
\end{align}
and the Majorana phase $\alpha_{21}$ and Dirac phase $\delta_{CP}$ are found in terms of the following relations:
\begin{align}
&
 \text{Im}[U^*_{e1} U_{e2}] = c_{12} s_{12} c_{13}^2 \sin \left( \frac{\alpha_{21}}{2} \right), \
 \text{Im}[U^*_{e1} U_{e3}] = - c_{12} s_{13} c_{13} \sin  \delta_{CP} 
,\\
&
 \text{Re}[U^*_{e1} U_{e2}] = c_{12} s_{12} c_{13}^2 \cos \left( \frac{\alpha_{21}}{2} \right), \
 \text{Re}[U^*_{e1} U_{e3}] = c_{12} s_{13} c_{13} \cos  \delta_{CP} 
,
\end{align}
where $\alpha_{21}/2,\ \delta_{CP}$
are subtracted from $\pi$, when $\cos(\alpha_{21}/2),\ \cos\delta_{CP}$ are negative.
In addition, the effective mass for the neutrinoless double beta decay is given by
\begin{align}
\langle m_{ee}\rangle=|\kappa||\tilde D_{\nu_1} \cos^2\theta_{12} \cos^2\theta_{13}+\tilde D_{\nu_2} \sin^2\theta_{12} \cos^2\theta_{13}e^{i\alpha_{21}}+\tilde D_{\nu_3} \sin^2\theta_{13}e^{-2i\delta_{CP}}|,
\end{align}
where its observed value could be measured by KamLAND-Zen in future~\cite{KamLAND-Zen:2016pfg}. 
%

\subsection{Radiative seesaw}
When $\eta_{1,2},\ \chi$ are inert bosons, the neutrino mass matrix is induced at one-loop level via mixings among neutral components of inert bosons.
Before discussing the neutrino sector, we formulate the Higgs sector.
The valid soft SUSY-breaking terms to construct the neutrino mass matrix are found as follows: 
\begin{align}
&-{\cal L}_{\rm soft} = \mu_{BH}^2 H_1 H_2  + \mu_{B\chi}^2 Y^{(6)}_1 \chi\chi +
A_a Y^{(4)}_1 H_1\eta_2 \chi+ A_b Y^{(4)}_1 H_2\eta_1 \chi
\nn\\
&
+m^2_{H_1}|H_1|^2+m^2_{H_2}|H_2|^2
+m^2_{\eta_1}|\eta_1|^2+m^2_{\eta_2}|\eta_2|^2+m^2_{\chi}|\chi|^2 + {\rm h.c.},\nn \label{eq:pot}
\end{align}
where  $m^2_{\eta_{1,2}}$, $m^2_{\chi}$
includes the invariant coefficients $1/(\tau^*-\tau)^{k_{\eta_{1,2},\chi}}$. 

 \subsubsection{Inert boson}
Inert bosons $\chi$, $\eta_1$, and $\eta_2$ mix each other through the soft {SUSY-breaking} terms of $A_{a,b}$ and $\mu_{B\eta}$, after the spontaneous electroweak symmetry breaking.
Here, we suppose to be $\mu_{B\eta},\ A_a<<A_b$ for simplicity, then the mixing dominantly comes from $\chi$ and $\eta_1$ only. 
This assumption does not affect the structure of the neutrino mass matrix.
Then the mass eigenstate is defined by
\begin{align}
\left[\begin{array}{c}
\chi_{R,I} \\ 
\eta_{1_{R,I}}  \\ 
\end{array}\right]=
\left[\begin{array}{cc}
c_{\theta_{R,I}} & -s_{\theta_{R,I}}  \\ 
s_{\theta_{R,I}} & c_{\theta_{R,I}}   \\ 
\end{array}\right]
\left[\begin{array}{c}
\xi_{1_{R,I}}   \\ 
\xi_{2_{R,I}}    \\ 
\end{array}\right],
\end{align}
 where $c_{\theta_{R,I}}, s_{\theta_{R,I}}$ are respectively the shorthand notations of $\sin\theta_{R,I}$ and $\cos\theta_{R,I}$,
 and $\xi_{1,2}$ are mass eigenstates for $\chi,\eta_1$ and their mass eigenvalues are denoted by $m_{i_{R,I}},\ (i=1,2)$. Notice that the mixing angle $\theta$ simultaneously diagonalizes the mass matrix of real and imaginary parts.

 \subsubsection{Mass matrix of $m_\nu$}
The active neutrino mass matrix $m_\nu$ is induced at one-loop level as follows:
\begin{align}
m_\nu &=- \frac{1}{2(4\pi)^2}
(y^T_\eta)_{i\alpha} (V_N)_{\alpha a} D_{N_a} (V_N^T)_{a\beta} (y_\eta)_{\beta j}\times
\nn\\
&\left[
s^2_{\theta_R} f(m_{\xi_{1_R}},D_{N_a}) 
+c^2_{\theta_R} f(m_{\xi_{2_R}},D_{N_a})
-s^2_{\theta_I} f(m_{\xi_{1_I}},D_{N_a})
-c^2_{\theta_I} f(m_{\xi_{2_I}},D_{N_a})
\right]
, \label{eq:mnu}\\
f(m_1,m_2)&=\int_0^1\ln\left[
x\left(\frac{m_1^2}{m_2^2}-1\right)+1
\right].
\end{align}
In order to fit the atmospheric mass square difference, we extract $\alpha_\eta$ from $y_\eta$ and redefine 
$m_\nu\equiv \alpha_\eta^2 \tilde m_\nu$.
Then, we can discuss the same manner as the case of canonical seesaw by regarding $\alpha_\eta^2$ as $\kappa$, where $\kappa$ is a parameter in the canonical seesaw model.

\section{Numerical analysis}
\label{sec:num}
In this section, we show numerical $\Delta \chi^2$ analysis for each of the cases, fitting the four reliable experimental data; $\Delta m_{\rm sol}^2, \sin^2\theta_{13},\sin^2\theta_{23},\sin^2\theta_{12}$ in ref.~\cite{Esteban:2020cvm}, where $\Delta m^2_{\rm atm}$ is supposed to be input value.~\footnote{We suppose CP phases $\delta_{\rm CP},\alpha_{21}$ to be predictive values, and the best fit values are applied for three charged-lepton masses.}
In case of IH for the radiative seesaw model, we would not find any allowed region within $5\sigma$. 
Thus, we do not discuss this case hereafter.
The dimensionful input parameters are randomly selected by the range of [$10^{2}-10^7$] GeV,
while the dimensionless ones [$10^{-10}-10^{-1}$] {except for $\tau$}.

\subsection{ NH for the canonical seesaw model} 
\begin{figure}[tb!]\begin{center}
\includegraphics[width=80mm]{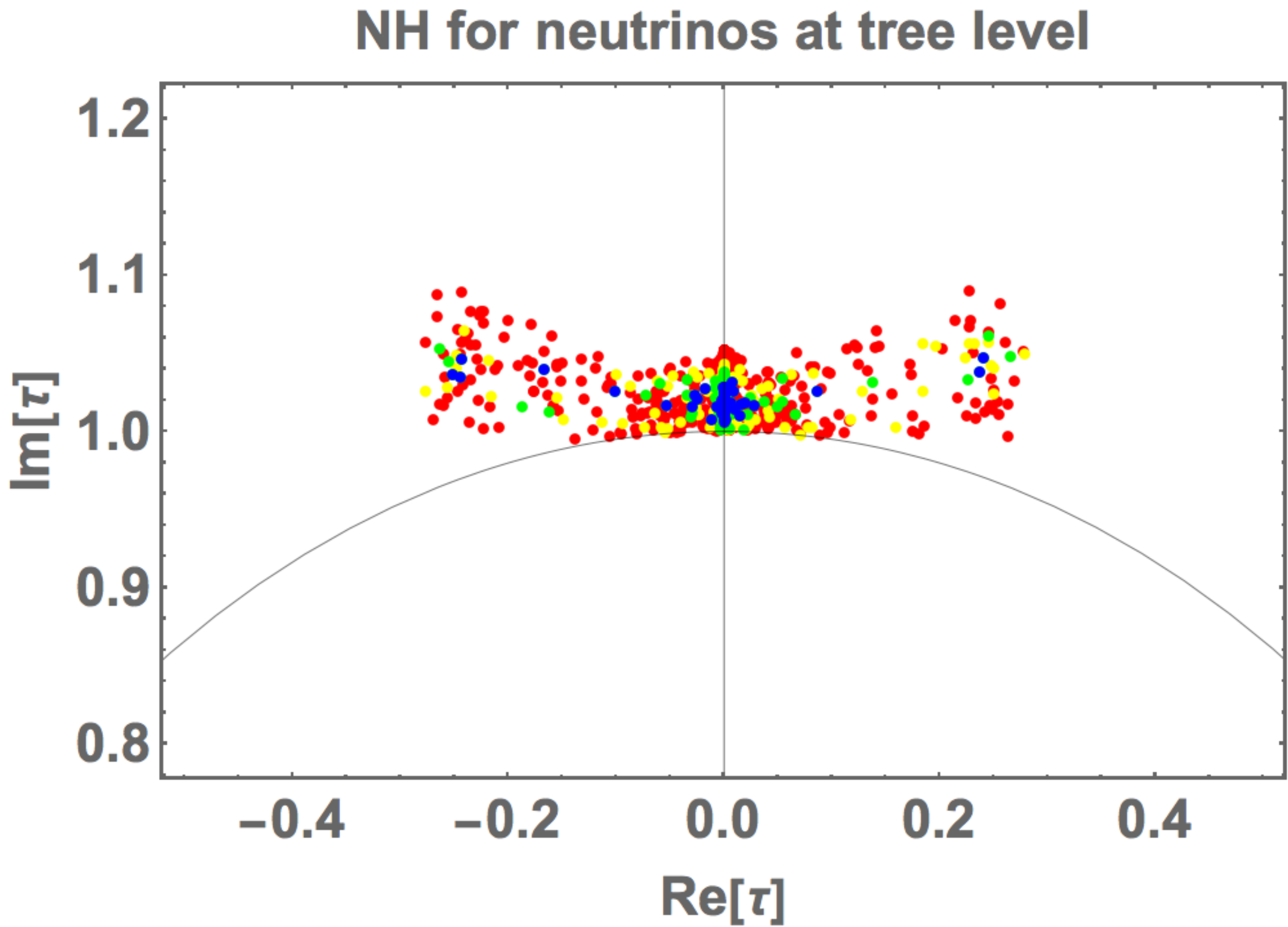} 
 \includegraphics[width=80mm]{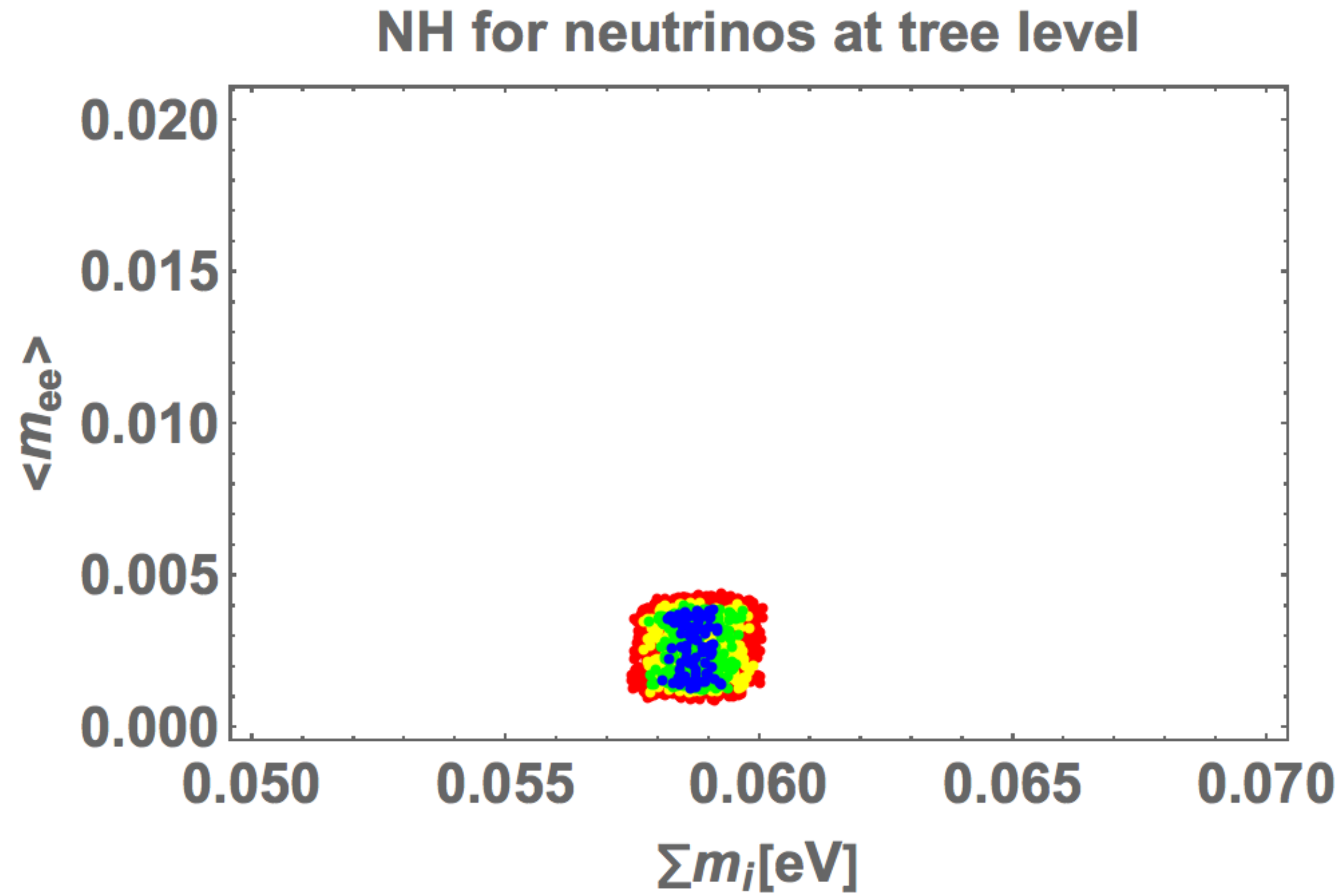}\\
  \includegraphics[width=80mm]{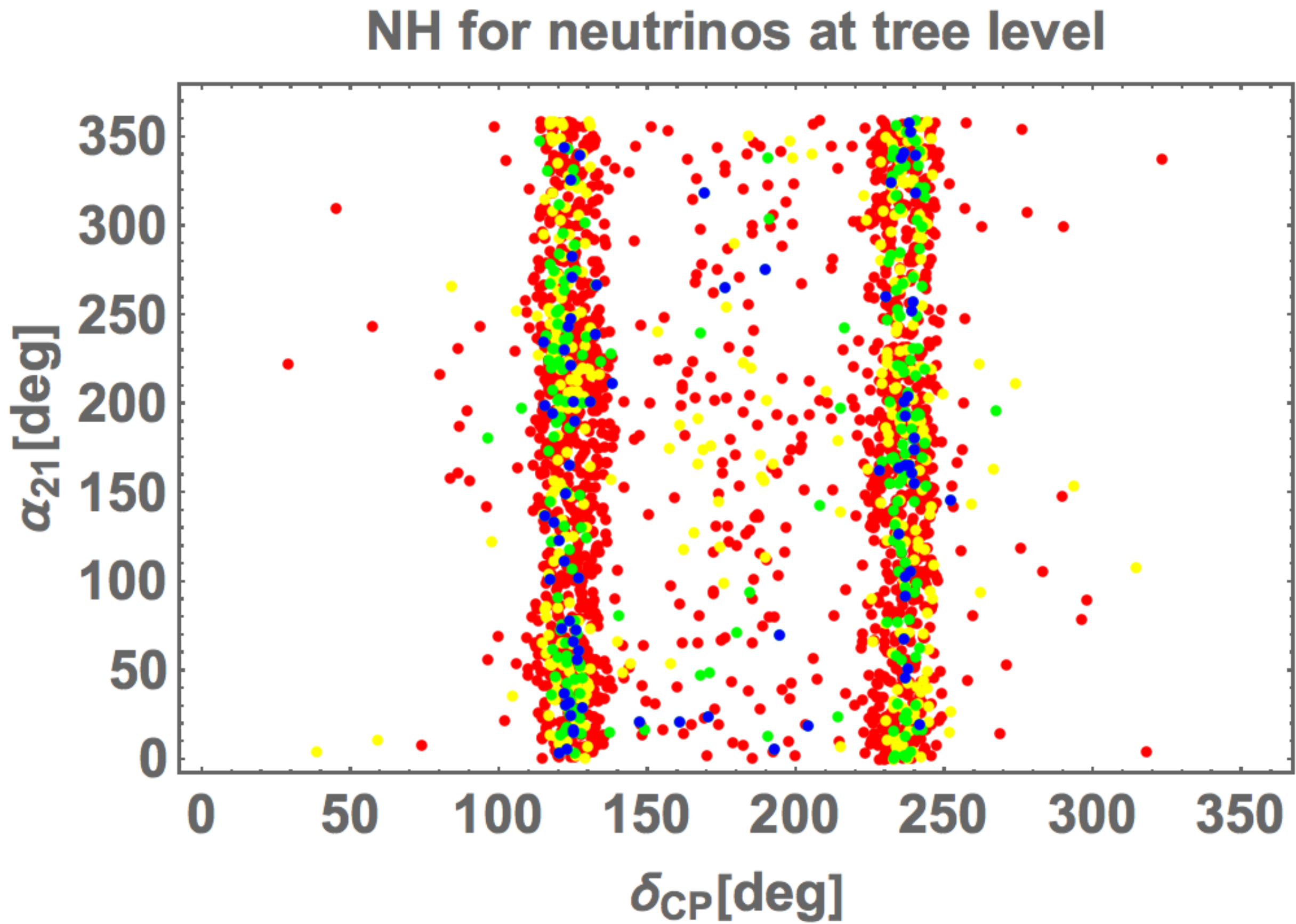}  
 \includegraphics[width=80mm]{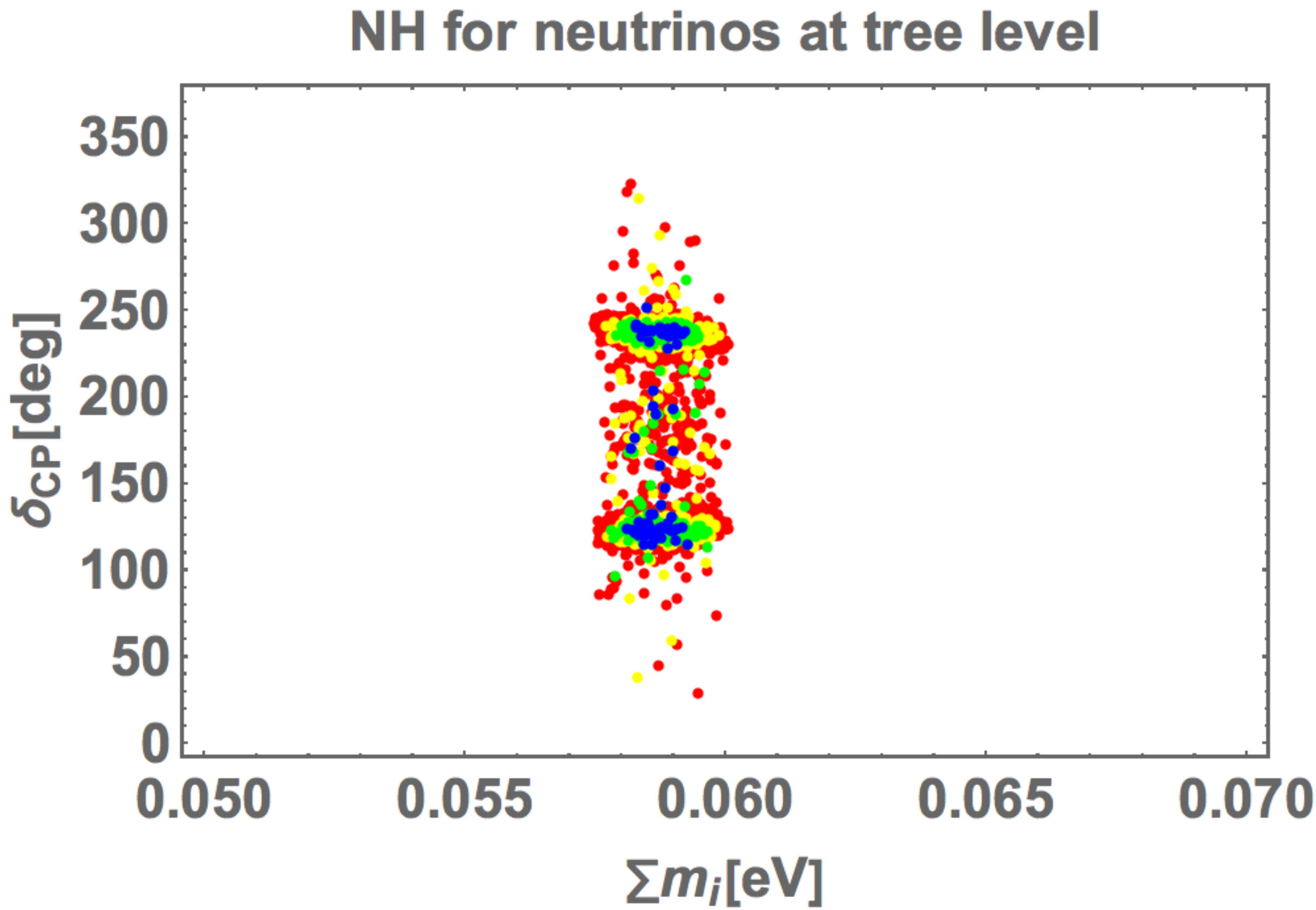}
\caption{
In case of NH for the canonical seesaw model, we show an allowed region of $\tau$ in the top left panel, $\langle m_{ee}\rangle$ in terms of sum of neutrino masses $\sum m_i$ in the top right one, Majorana phase $\alpha_{21}$and Dirac CP phase $\delta_{\rm CP}$ in the bottom left one, and Dirac CP phase $\delta_{\rm CP}$ versus sum of neutrino masses $\sum m_i$ in the bottom right one, respectively.
Each of colors corresponds to the range of $\Delta\chi^2$ value such that blue: $\Delta\chi^2 \leq 1$, green: $1< \Delta\chi^2 \le 4$, yellow: $4< \Delta\chi^2 \le 9$, and red: $9< \Delta\chi^2 \le 25$.}
\label{fig:nh-tree}
\end{center}\end{figure}

Fig.~\ref{fig:nh-tree} represents NH for the canonical seesaw model. 
We show an allowed region of $\tau$ in the top left panel, $\langle m_{ee}\rangle$ in terms of sum of neutrino masses $\sum m_i$ in the top right one, Majorana phase $\alpha_{21}$ and Dirac CP phase $\delta_{\rm CP}$ in the bottom left one, and Dirac CP phase $\delta_{\rm CP}$ versus sum of neutrino masses $\sum m_i$ in the bottom right one, respectively.
Each of colors corresponds to the range of $\Delta\chi^2$ value such that blue: $\Delta\chi^2 \leq 1$, green: $1< \Delta\chi^2 \le 4$, yellow: $4< \Delta\chi^2\le 9$, and red: $9< \Delta\chi^2 \le 25$.
These figures suggest, within $5\sigma$, that $0.058{\rm eV}\lesssim\sum m_i\lesssim0.06$ eV,
$0.001{\rm eV}\lesssim\langle m_{ee}\rangle\lesssim0.004$ eV, any value is possible for $\alpha_{21}$, and $\delta_{CP}$ tends to be localized at nearby 120$^\circ$ and 240$^\circ$.

\subsection{ IH for the canonical seesaw model} 
\begin{figure}[tb!]\begin{center}
\includegraphics[width=80mm]{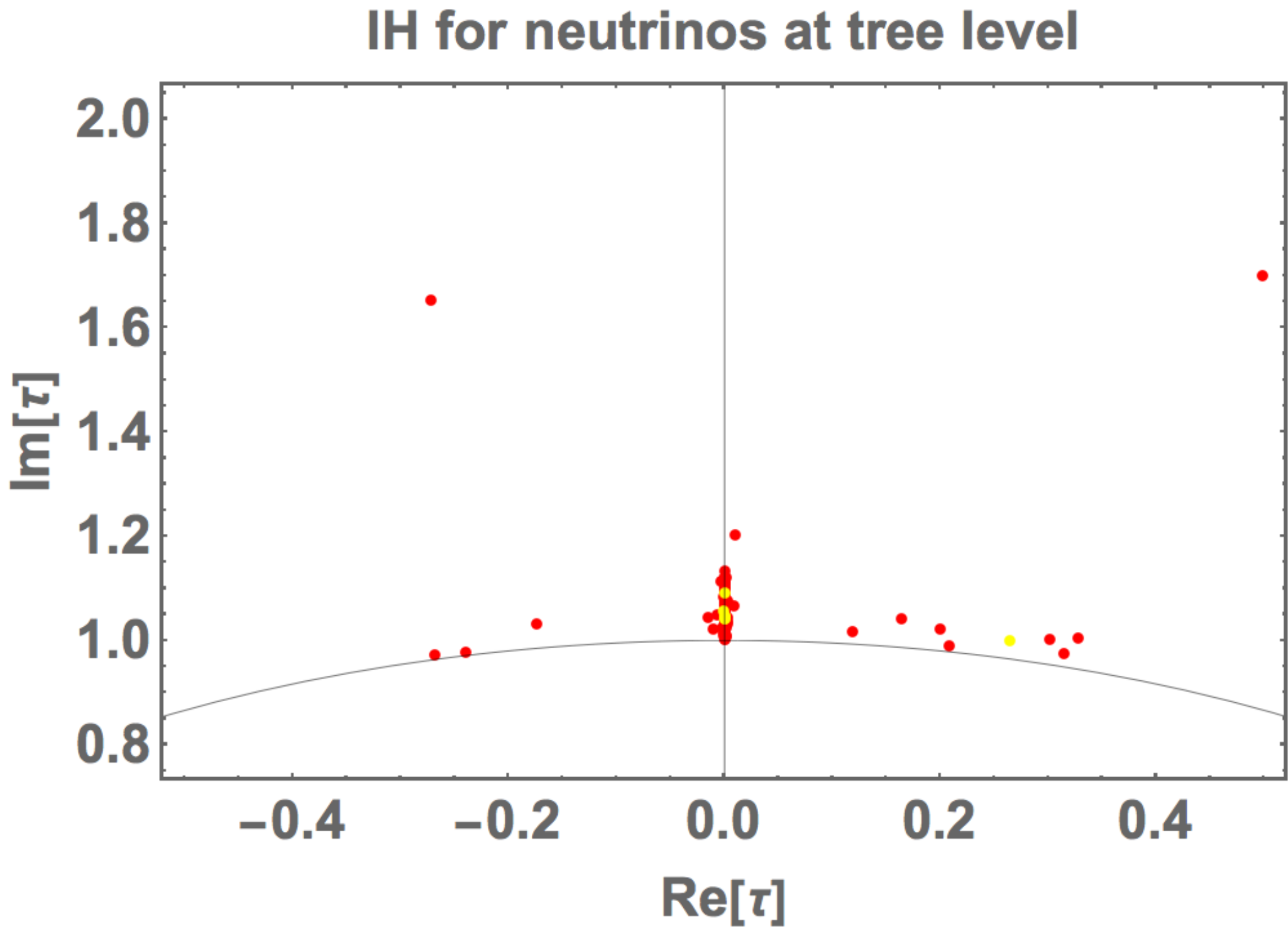} 
 \includegraphics[width=80mm]{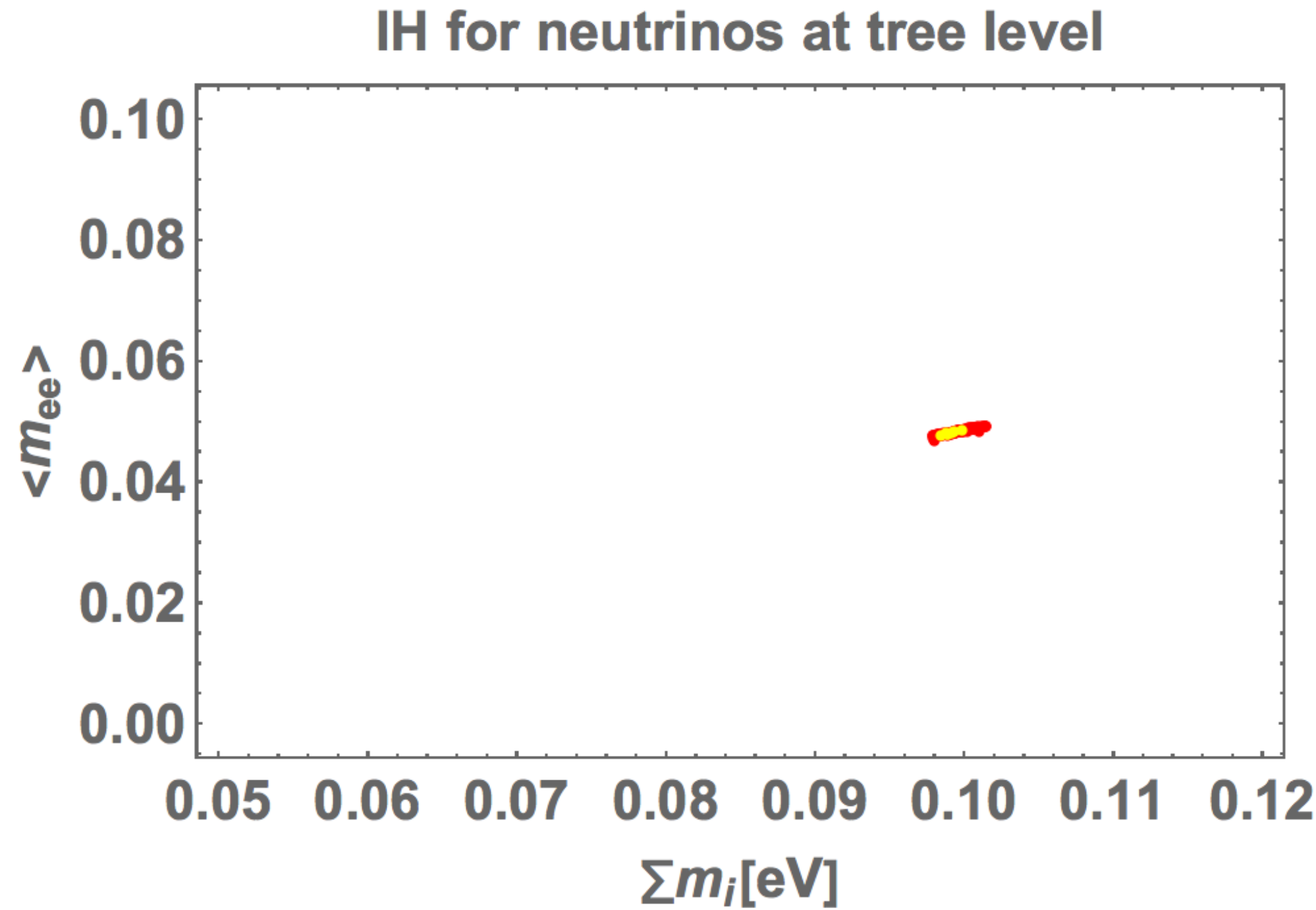}\\
  \includegraphics[width=80mm]{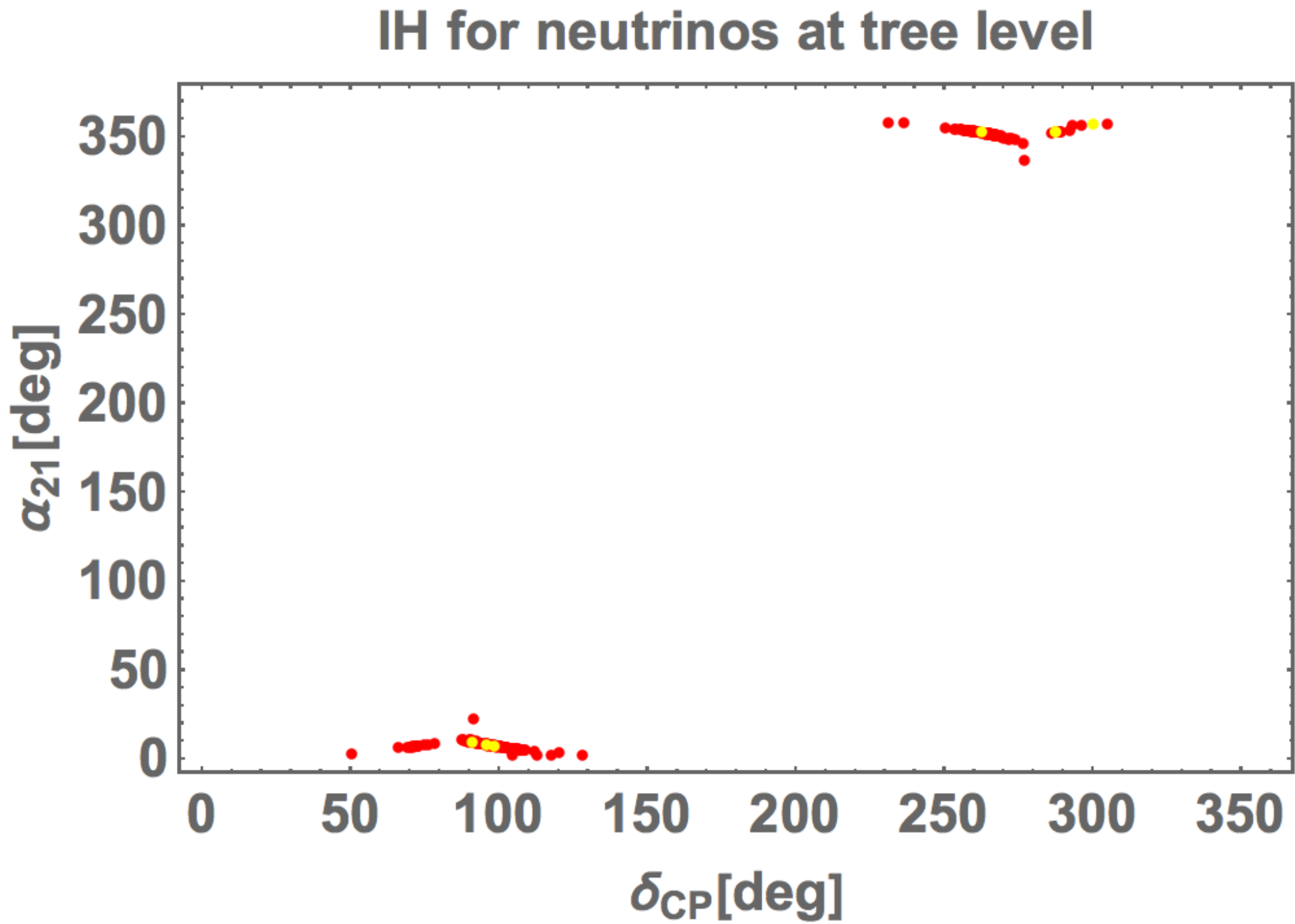}  
 \includegraphics[width=80mm]{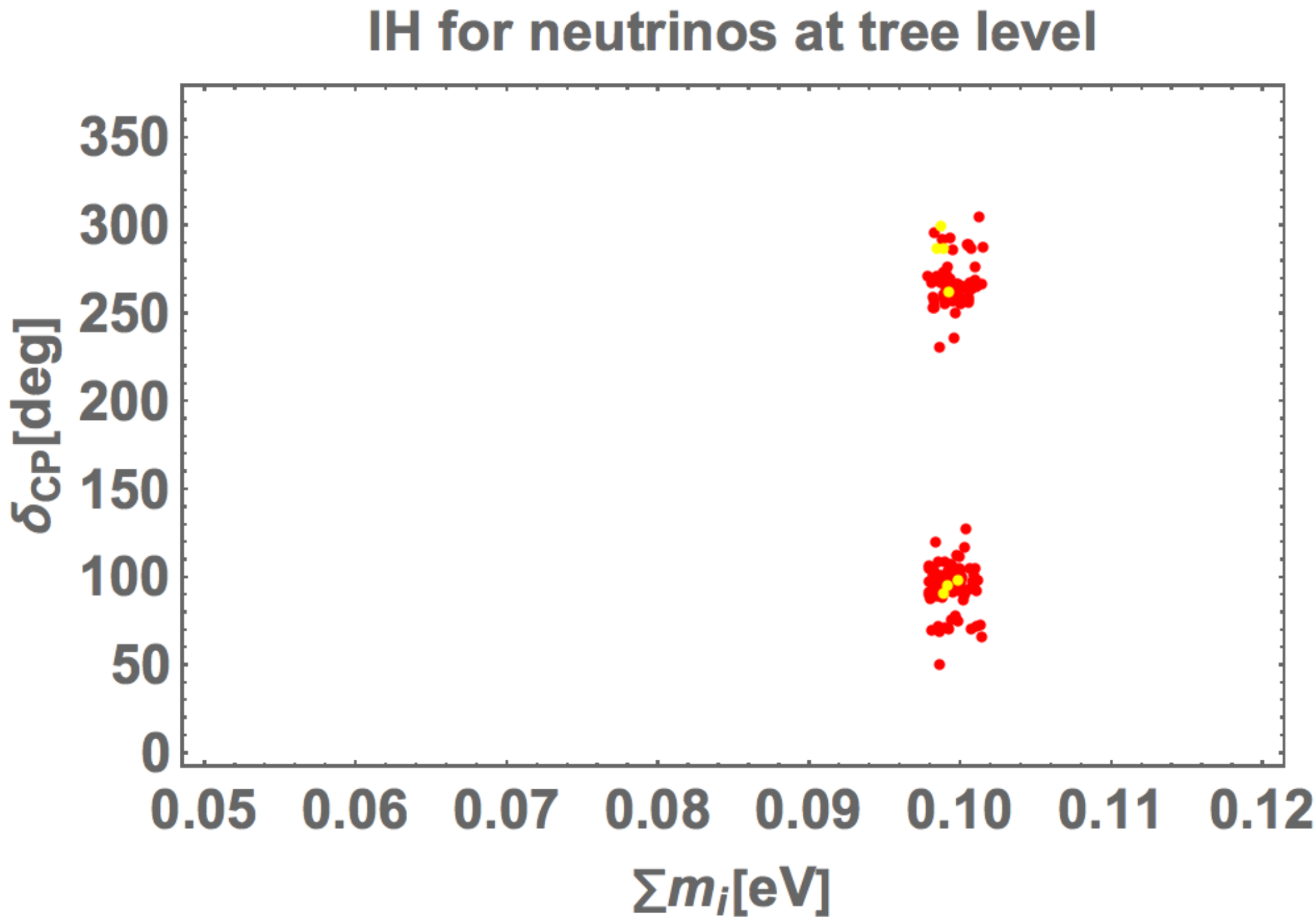}
 \caption{
In case of IH for the canonical seesaw model, where the legends and the colors are the same as the NH for canonical seesaw case.}
\label{fig:ih-tree}
\end{center}\end{figure}

Fig.~\ref{fig:ih-tree} represents IH for the canonical seesaw model, where the legends and the colors are the same as the NH for canonical seesaw case.
These figures suggest that $0.098\ {\rm eV}\lesssim\sum m_i\lesssim0.102\ {\rm eV}$,
$\langle m_{ee}\rangle \simeq 0.05$ eV, $\alpha_{21} \simeq 0^\circ$, and $50^\circ\lesssim\delta_{CP}\lesssim130^\circ$ and $230^\circ\lesssim\delta_{CP}\lesssim310^\circ$.

\subsection{ NH for the radiative seesaw model} 
\begin{figure}[tb!]\begin{center}
\includegraphics[width=80mm]{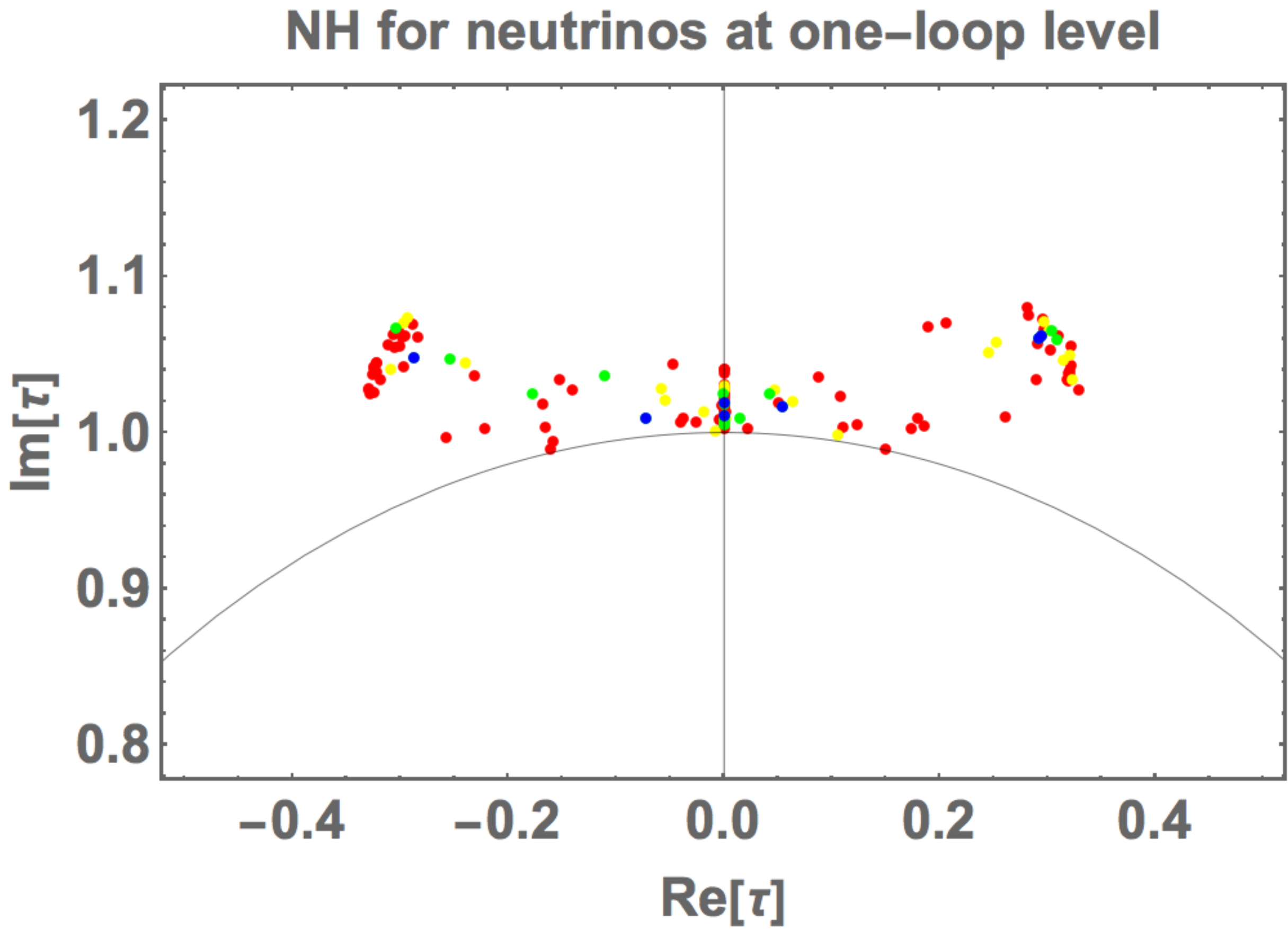} 
 \includegraphics[width=80mm]{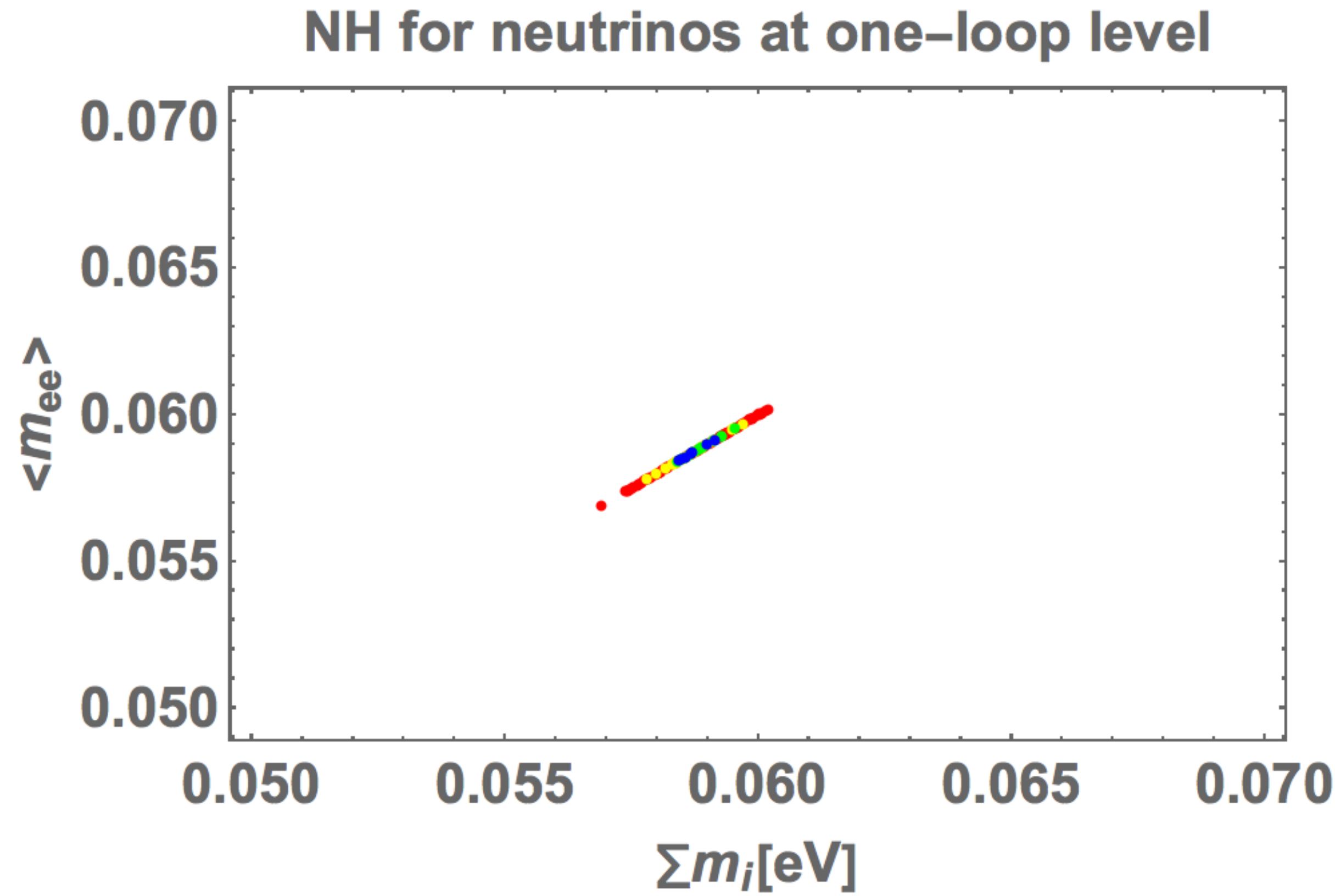}\\
  \includegraphics[width=80mm]{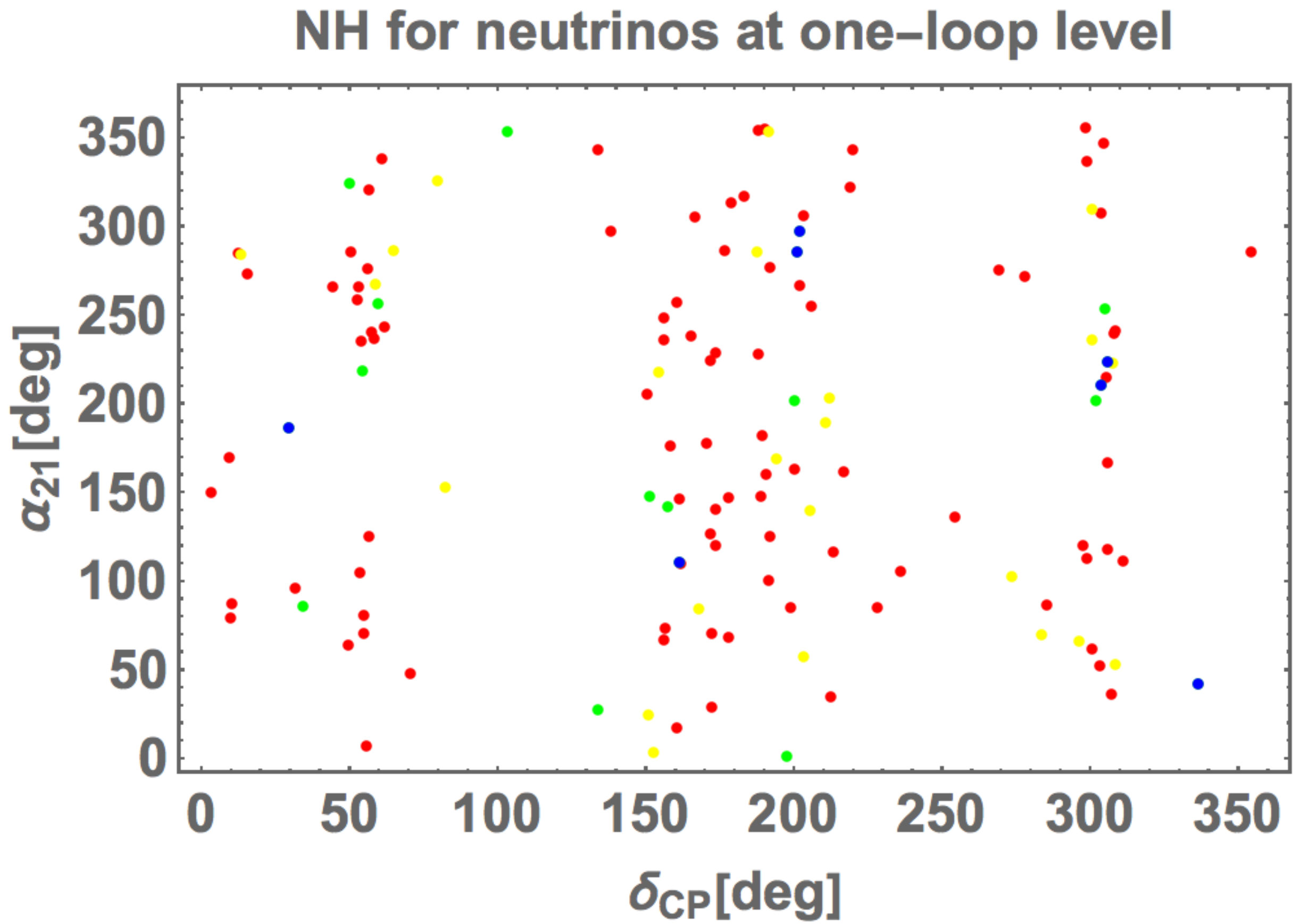}  
 \includegraphics[width=80mm]{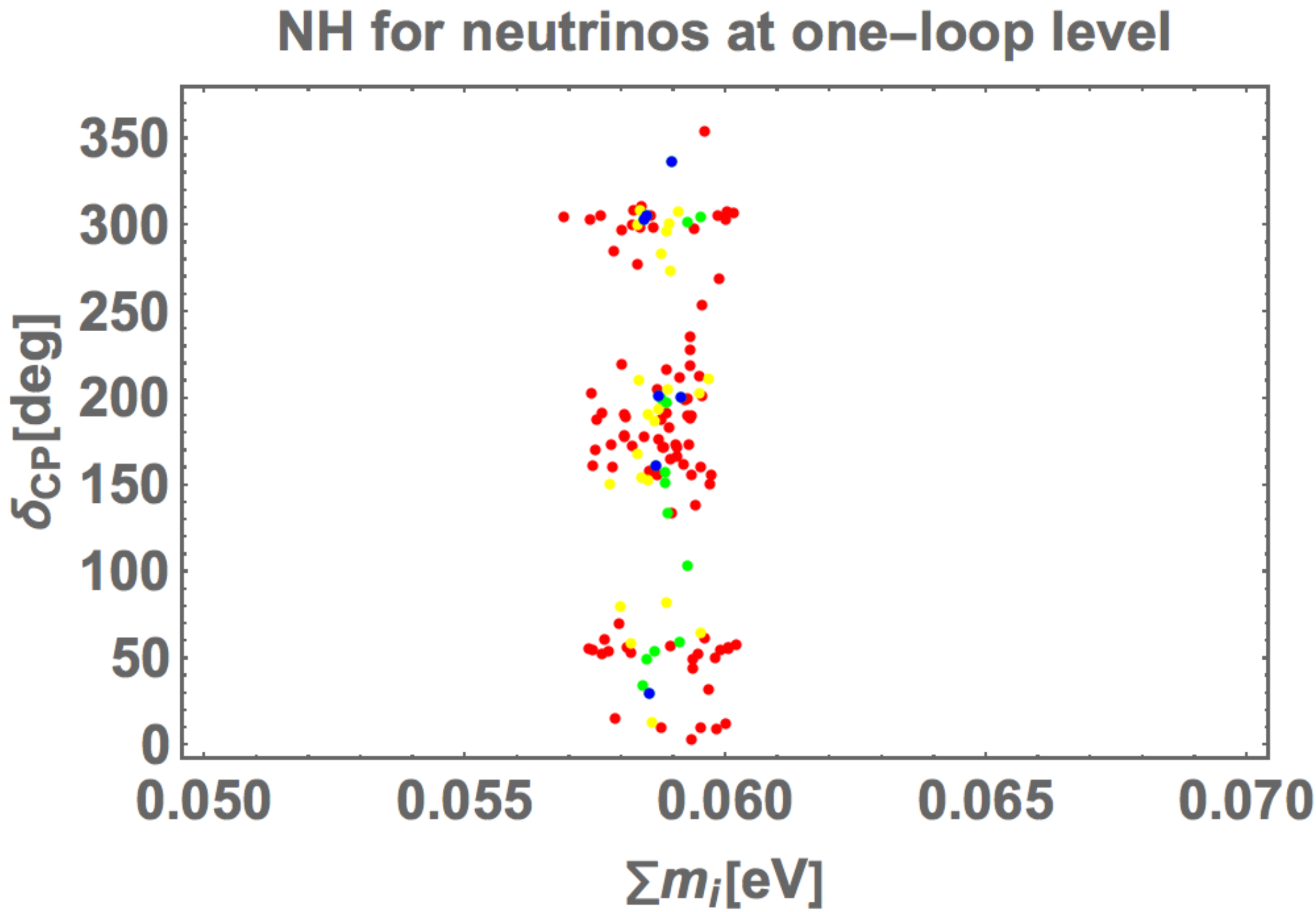}
 \caption{
In case of NH for the radiative seesaw model, where the legends and the colors are the same as the NH for canonical seesaw case.}
\label{fig:nh-rad}
\end{center}\end{figure}

Fig.~\ref{fig:nh-rad} represents NH for the radiative seesaw model, where the legends and the colors are the same as the NH for canonical seesaw case.
These figures suggest that $0.057{\rm eV}\le(\sum m_i,\langle m_{ee}\rangle)\le0.06$ eV, and any values are allowed for phases.

\section{Conclusion and discussion}
\label{sec:conclusion}

We have studied a double covering of modular $A_4$ flavor symmetry
in which we have constructed lepton models in cases of canonical seesaw and radiative seesaw models. Then, we have some predictions for both the cases except for IH of radiative seesaw.

\section*{Acknowledgments}
The work of H. O. is supported by the Junior Research Group (JRG) Program at the Asia-Pacific Center for Theoretical
Physics (APCTP) through the Science and Technology Promotion Fund and Lottery Fund of the Korean Government and was supported by the Korean Local Governments-Gyeongsangbuk-do Province and Pohang City. 
H. O. is sincerely grateful for all the KIAS members.
Y. O. was supported from European Regional Development Fund-Project Engineering Applications of Microworld
Physics (No.CZ.02.1.01/0.0/0.0/16\_019/0000766)

{

 \appendix

\section{Formulas in modular $T^\prime$ framework}

In this appendix, we summarize some formulas in the framework of $T^\prime$ modular symmetry belonging to 
the $SL(2,\mathbb{Z})$ modular symmetry. 
The $SL(2,Z_3)$ modular symmetry corresponds to the $T^\prime$ modular symmetry. 
The modulus $\tau$ transforms as
\begin{align}
& \tau \longrightarrow \gamma\tau= \frac{a\tau + b}{c \tau + d},
\end{align}
with $\{a,b,c,d\} \in Z_3$ satisfying $ad-bc=1$ and ${\rm Im} [\tau]>0$.
The transformation of modular forms $f(\tau)$ are given by
\begin{align}
& f(\gamma\tau)= (c\tau+d)^k f(\tau)~, ~~ \gamma \in SL(2,Z_3)~ ,
\end{align}
where $f(\tau)$ denotes holomorphic functions of $\tau$ with the modular weight $k$.

In a similar way, the modular transformation of a matter chiral superfield $\phi^{(I)}$ with the modular weight $-k_I$ 
is given by 
\begin{equation}
\phi^{(I)} \to (c\tau+d)^{-k_I}\rho^{(I)}(\gamma)\phi^{(I)},
\end{equation}
where $\rho^{(I)}(\gamma)$ stands for an unitary matrix corresponding to $T^\prime$ transformation. 
Note that the superpotential is invariant when the sum of modular weight from fields and modular form is zero and the term is a singlet under the $T^\prime$ symmetry. 
It restricts a form of the superpotential as shown in Eq. (\ref{eq:sp-lep}).

Modular forms are constructed on the basis of weight 1 modular form, $ Y^{(1)}_2=(Y_1, Y_2)^T$, transforming
as a doublet of $T^\prime$. 
Their explicit forms are written by the Dedekind eta-function $\eta(\tau)$ with respect to $\tau$ \cite{Feruglio:2017spp, Liu:2019khw}:
\begin{eqnarray} 
\label{eq:Y-A4}
Y_{1}(\tau) &=& \sqrt{2}e^{i\frac{7\pi}{12}} \frac{\eta^3(3\tau)}{\eta(\tau)}, \nonumber \\
Y_{2}(\tau) &=&  \frac{\eta^3(3\tau) + 1/3\eta^3(\tau/3)}{\eta(\tau)}. \label{eq:Yi}
\nonumber
\end{eqnarray}
%
Modular forms of higher weight can be obtained from tensor products of $Y^{(1)}_2$. 
We enumerate some modular forms used in our analysis:
\begin{align}
Y_1^{(4)} 
&=  
-4 Y_1^3 Y_2 - (1-i) Y_2^4, 
\\
Y^{(6)}_{\bf 1}
&=
(1-i)e^{i\pi/6}Y_2^6
-(1+i)e^{i\pi/6}Y_1^6
- 10 e^{i\pi/6}Y_1^3Y_2,
\\
Y^{(2)}_3
&\equiv(f_1,f_2,f_3)^T
=
(
e^{i\pi/6}Y_2^2, 
\sqrt{2} e^{i7\pi/12} Y_1 Y_2, 
Y_1^2
)^T, 
\\
Y^{(3)}_{2} 
&\equiv(y_1,y_2)^T
=
(
3e^{i\pi/6}Y_1 Y_2^2, 
\sqrt{2}e^{i5\pi/12}Y_1^3
- e^{i\pi/6}Y_2^3
)^T, 
\\
Y^{(3)}_{2^{\prime\prime}} 
&\equiv
(y^{\prime\prime}_1,
y^{\prime\prime}_2)^T
=
(
Y_1^3 + (1-i) Y_2^3,
-3 Y_1^2 Y_2
)^T.
\end{align}


}


\end{document}